# Intelligent energy management of steam generators

Ahmed S. Hussein
Electrical Power and Machines Department
Aswan University, Aswan, Egypt
eng.ahmeds09@gmail.com

Noha H. El-Amary
Electrical and Control Department
Arab Academy for Science, Technology and Maritime Transport (AASTMT),
Cairo, Egypt

Loai Saad El-din Nasrat
Electrical Power and Machines Department,
Aswan University,
Aswan, Egypt

Ali Selim
Electrical Power and Machines Department, Aswan University,
Aswan, Egypt

***Abstract*— This paper introduces a smart model for intelligent energy management of steam generators which are utilized for steam generator and controlling the air to fuel ratio for steam generator all over the firing curve and transient mode operation. Nowadays, the environment faces a lot of pollution and global warming phenomena. With the spread of electrical devices, electric cars with conventional electrical generation sources, and the increase in electrical consumption, instead of minimizing the pollution level the situation becomes disastrous. Steam generators have a lot of pros which cannot be neglected, such as: high efficiency, reliable operation, low emission (with regular maintenance), and big variety of fuel source. However, regular maintenance overlooks some parameters, especially the air to fuel ratio that achieves green environment, high efficiency and low fuel consumption. The steam generator system is simulated utilizing Simulink/MATLAB. The system is operated at different loading and generation conditions to determine the variation of air to fuel ratio against power variation. Neural Network (NN) unit is added in different locations and scenarios. It is effective in controlling the main bus of air, fuel, auxiliary and inverter speed. By testing the NN on the simulated tested system, the results are satisfied.***

## 1. INTRODUCTION

In industrial furnaces, the air-fuel ratio is the ratio of the mass of air to the mass of fuel in the fuel-air mixture. The stoichiometric air-fuel ratio is the ratio at which the fuel is completely burned with no excess air or fuel. For most industrial fuels, the stoichiometric air-fuel ratio is approximately 14.7:1. [3]

Imagine crucially needing to create the perfect recipe, but instead of flour, sugar, and spices, your ingredients are air and fuel. In the world of internal combustion engines, that's exactly what engineers grapple with – achieving the ideal air-fuel ratio. This ratio, expressed as a mass of air to a mass of fuel, plays a vital role in how efficiently and cleanly an engine operates.

Just like in baking, a precise balance is key. Too much flour and your cake end up dense; too little, and it crumbles. Similarly, an engine's performance suffers with an improper air-fuel mixture. The ideal ratio, known as the stoichiometric ratio, ensures complete combustion of the fuel, maximizing power output and minimizing harmful emissions.

But why is this ratio so critical? Well, internal combustion engines rely on burning fuel to create energy. That fuel needs oxygen, which is primarily obtained from the air that enters the engine. If there's not enough air (lean mixture), the fuel doesn't burn completely, leading to wasted fuel, reduced power, and increased emissions. Conversely, too much air (rich mixture) can also cause problems, like overheating and higher hydrocarbon emissions.

Finding the perfect balance between power, efficiency, and emissions is an ongoing quest for engine designers. Modern engines utilize sophisticated engine management systems that constantly monitor and adjust the air-fuel ratio based on various factors like engine speed, load, and temperature.

Understanding the air-fuel ratio is not just for engineers. It's a fundamental concept for anyone interested in how internal combustion engines work. By delving deeper into this concept, we can appreciate the intricate dance between air and fuel that propels our cars, trucks, and other machinery.

In the following sections, we'll explore the science behind the air-fuel ratio, delve into the concept of lean and rich mixtures, and discover how engine management systems maintain the optimal balance for peak performance and reduced environmental impact.

## 2. DIFFERENT FURNACES FUELS

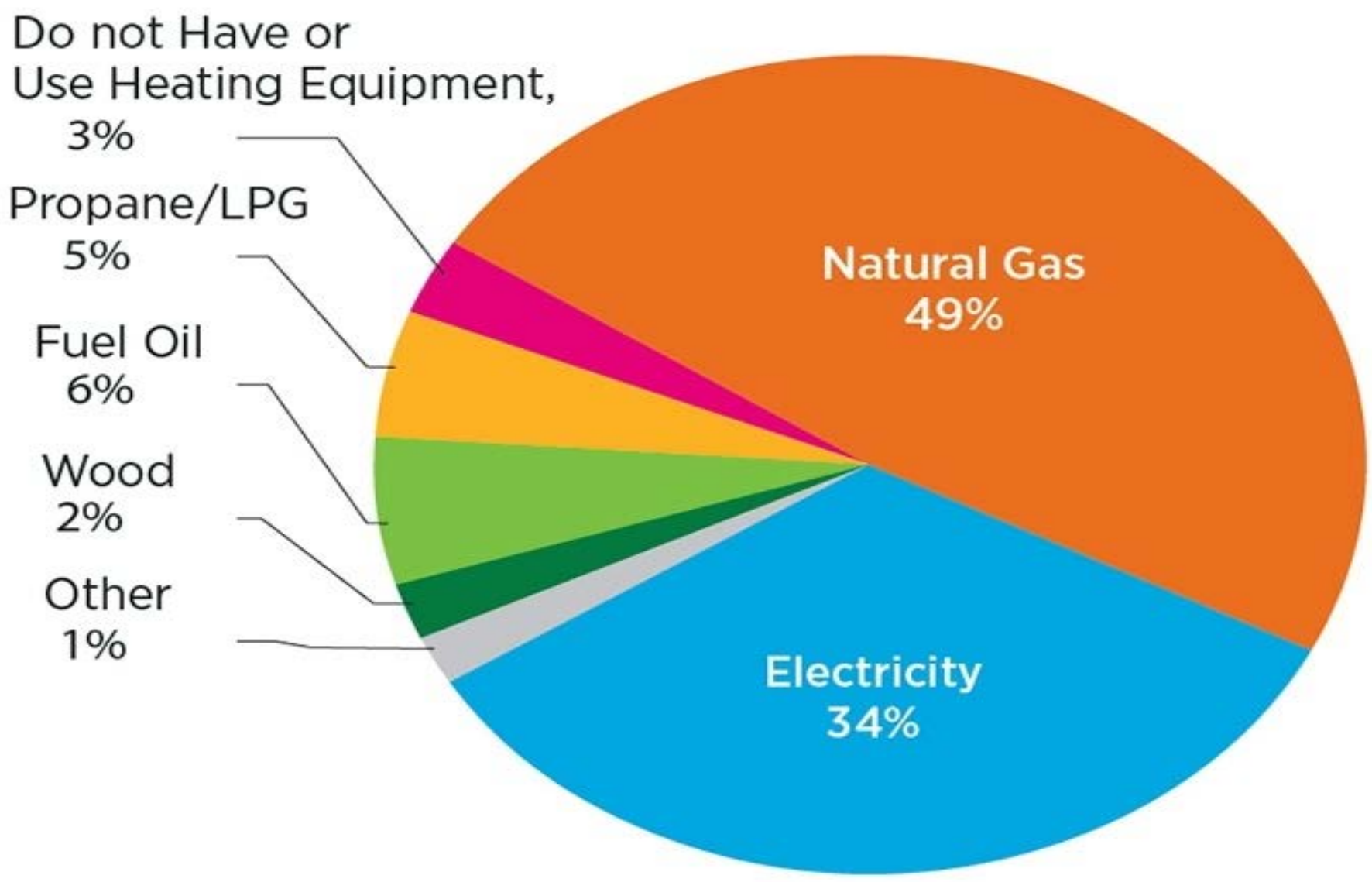

Furnaces are used to heat materials to high temperatures for a variety of purposes, including melting, annealing, and brazing. The type of fuel used in a furnace depends on the temperature required, the desired atmosphere, and the cost of the fuel.

2.1 Natural gas: Natural gas is a clean-burning fuel that is relatively inexpensive. It is often used in residential and commercia furnaces. [4]

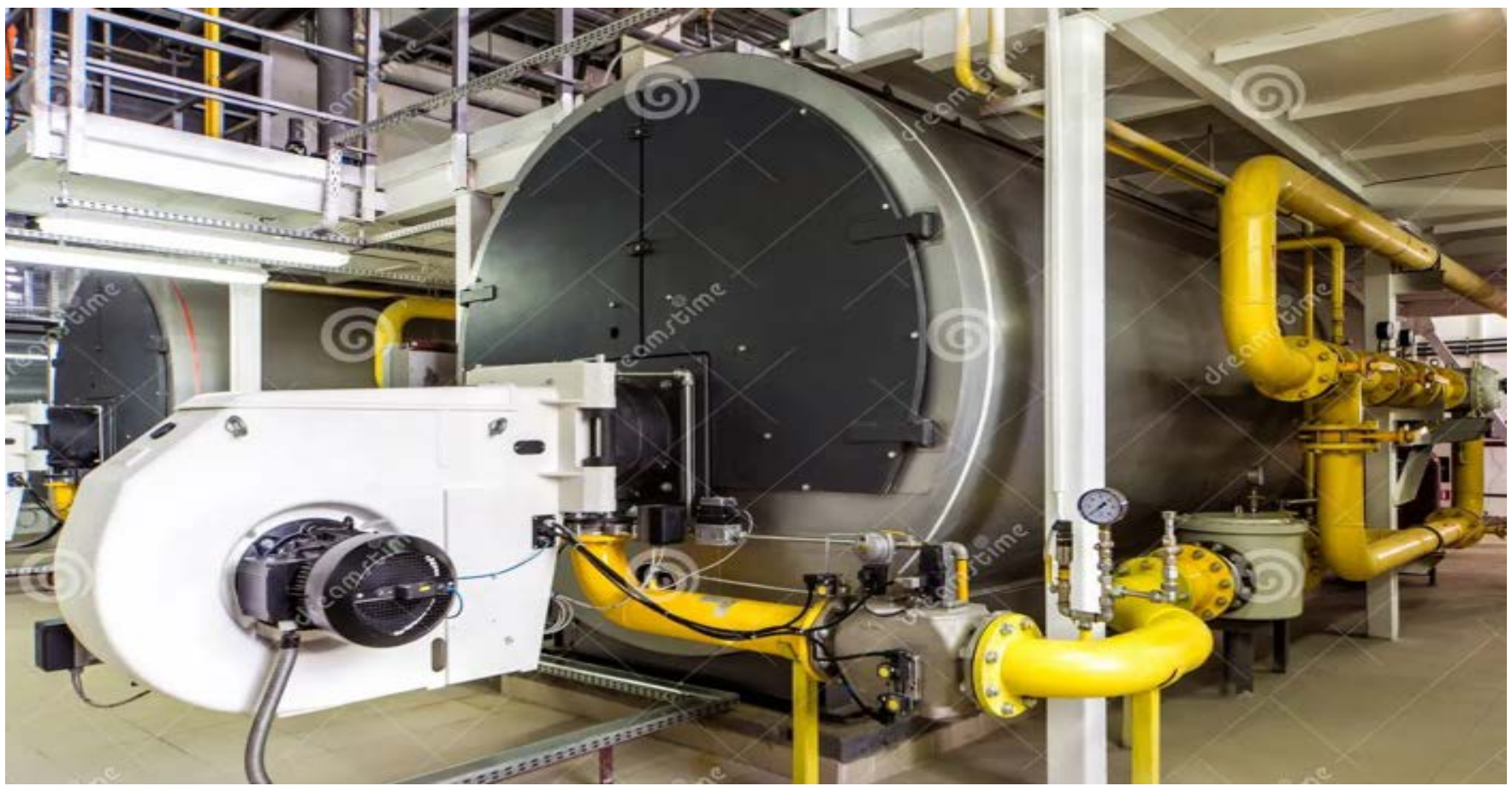

2.2 Propane: Propane is a gas that is similar to natural gas. It is also a clean-burning fuel, but it is more expensive than natural gas. Propane is often used in industrial furnaces. [4]

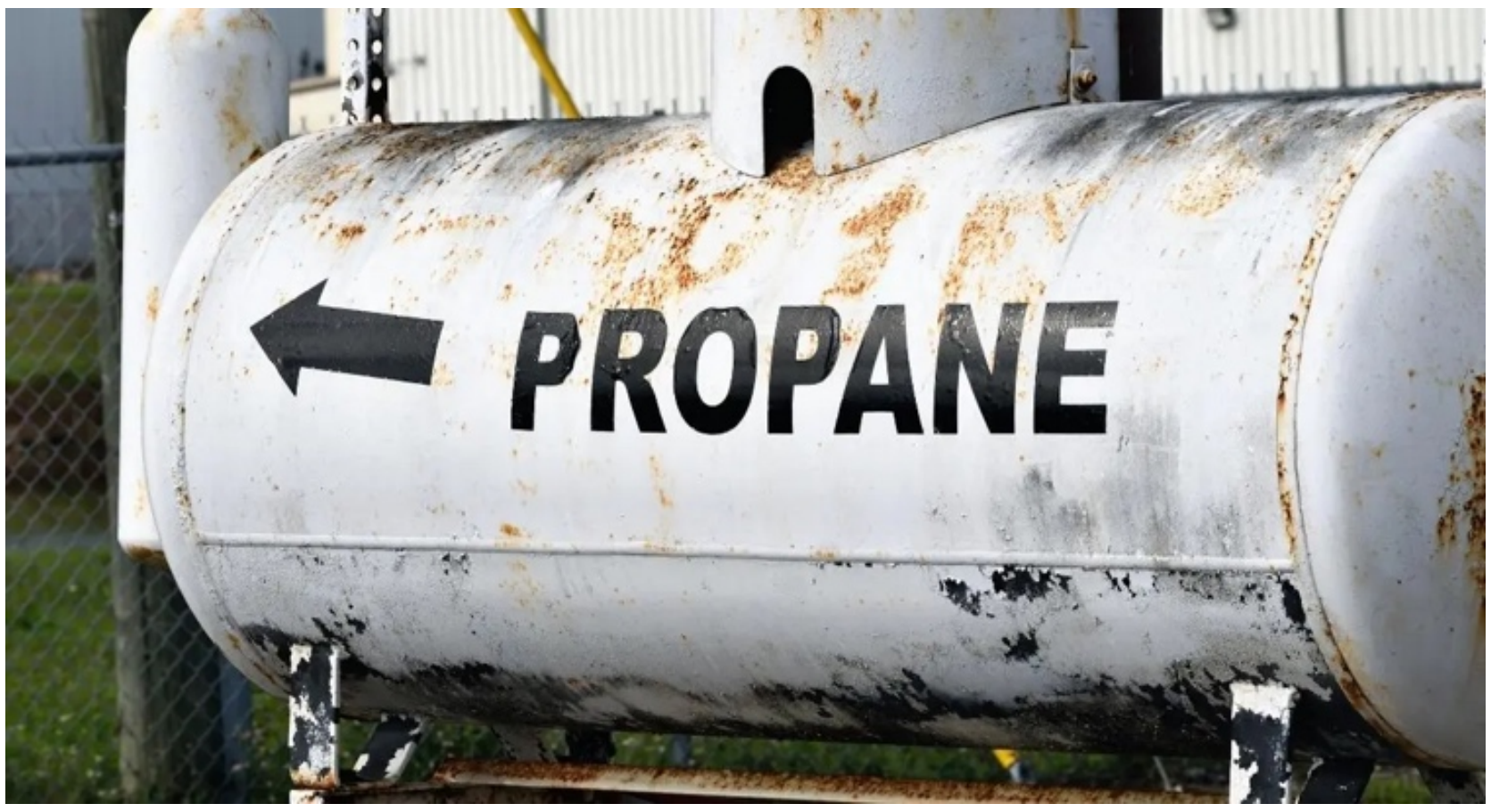

2.3 Oil: Oil is a fossil fuel that is burned to produce heat. It is less efficient than natural gas or propane, but it is still a popular choice for industrial furnaces.

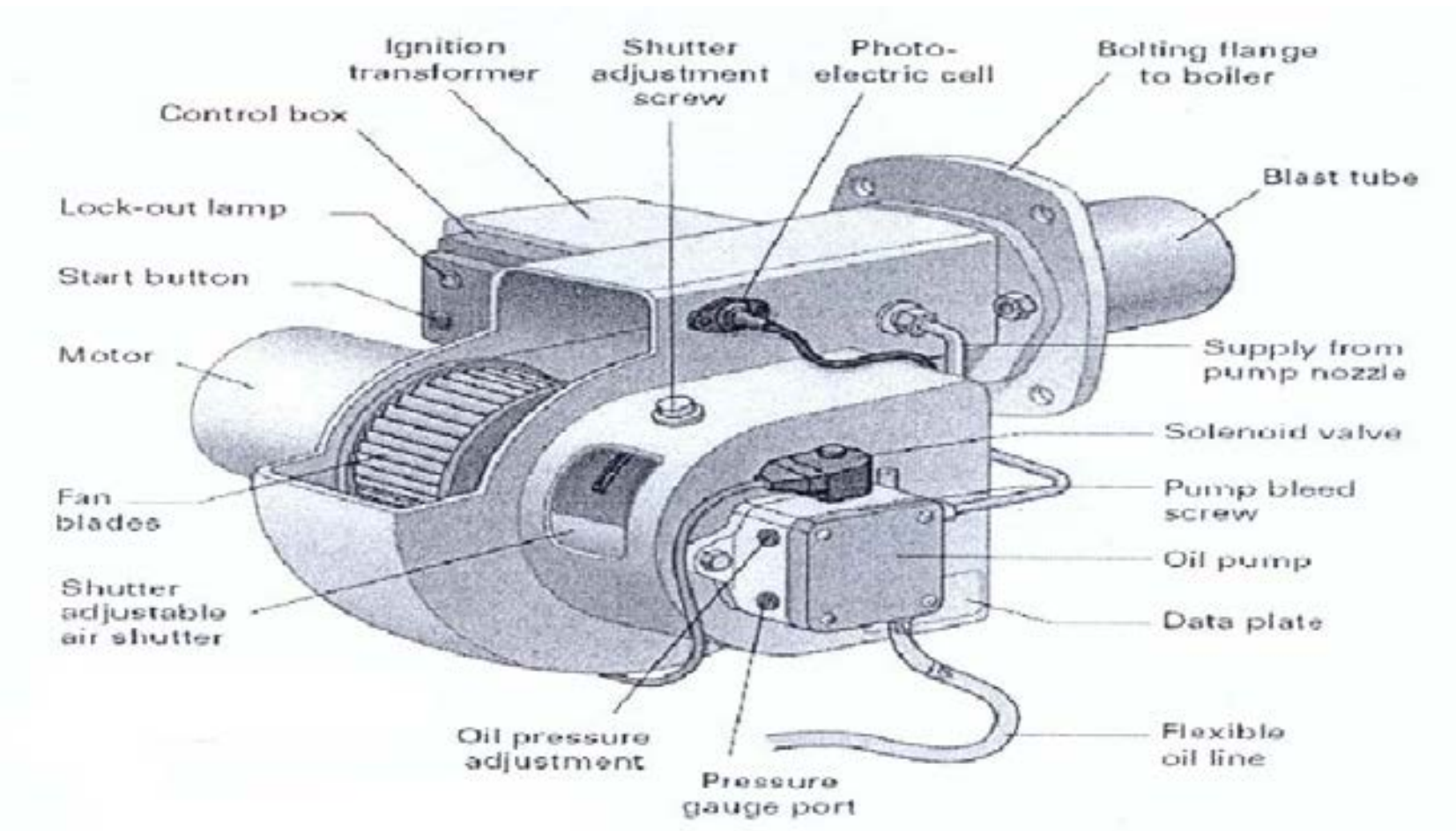

2.4 Electricity: Electric furnaces use electricity to heat elements that then heat the air or other material in the furnace. They are very efficient, but they can be expensive to operate. [4]

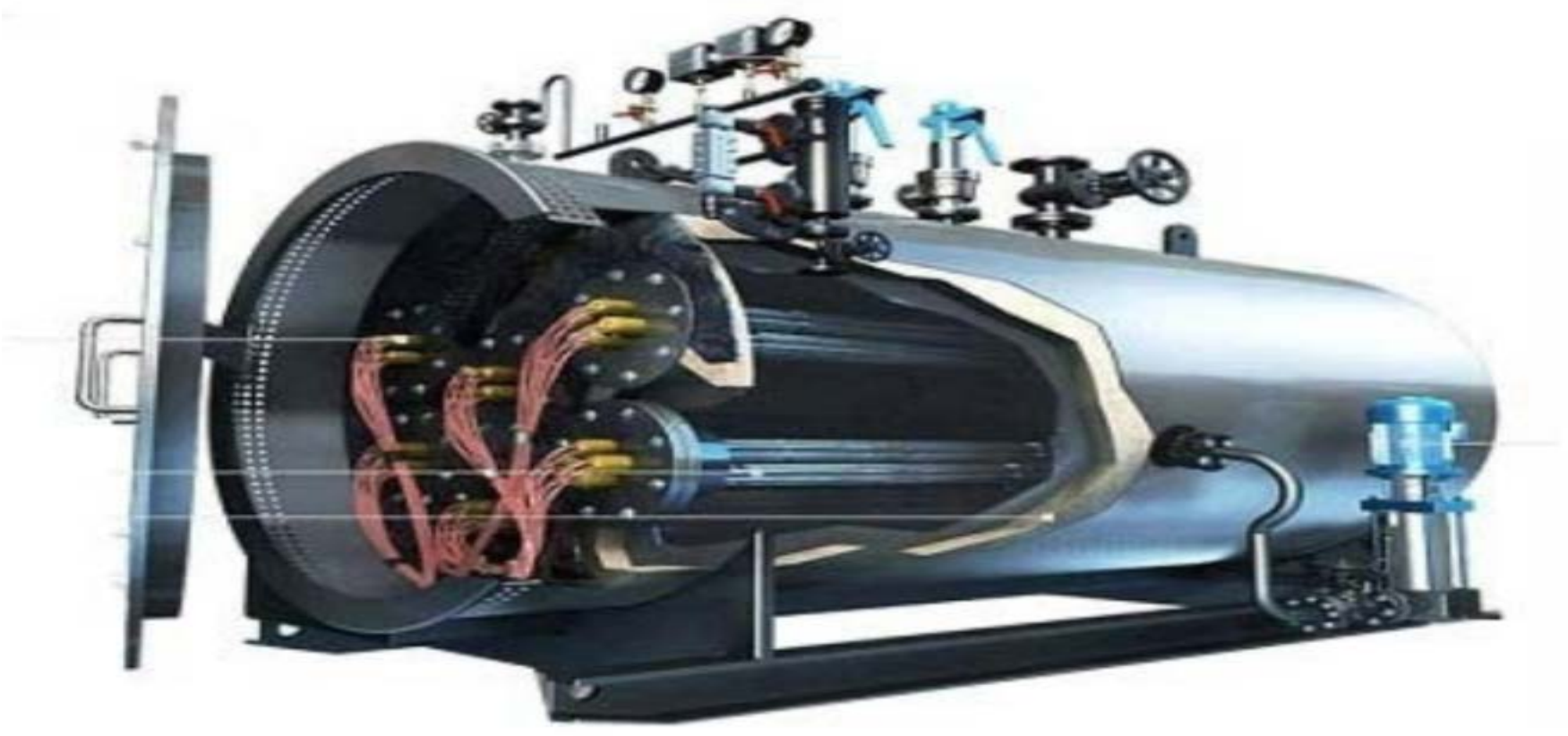

2.5 Wood: Wood is a renewable fuel that can be burned to produce heat. It is often used in small furnaces.

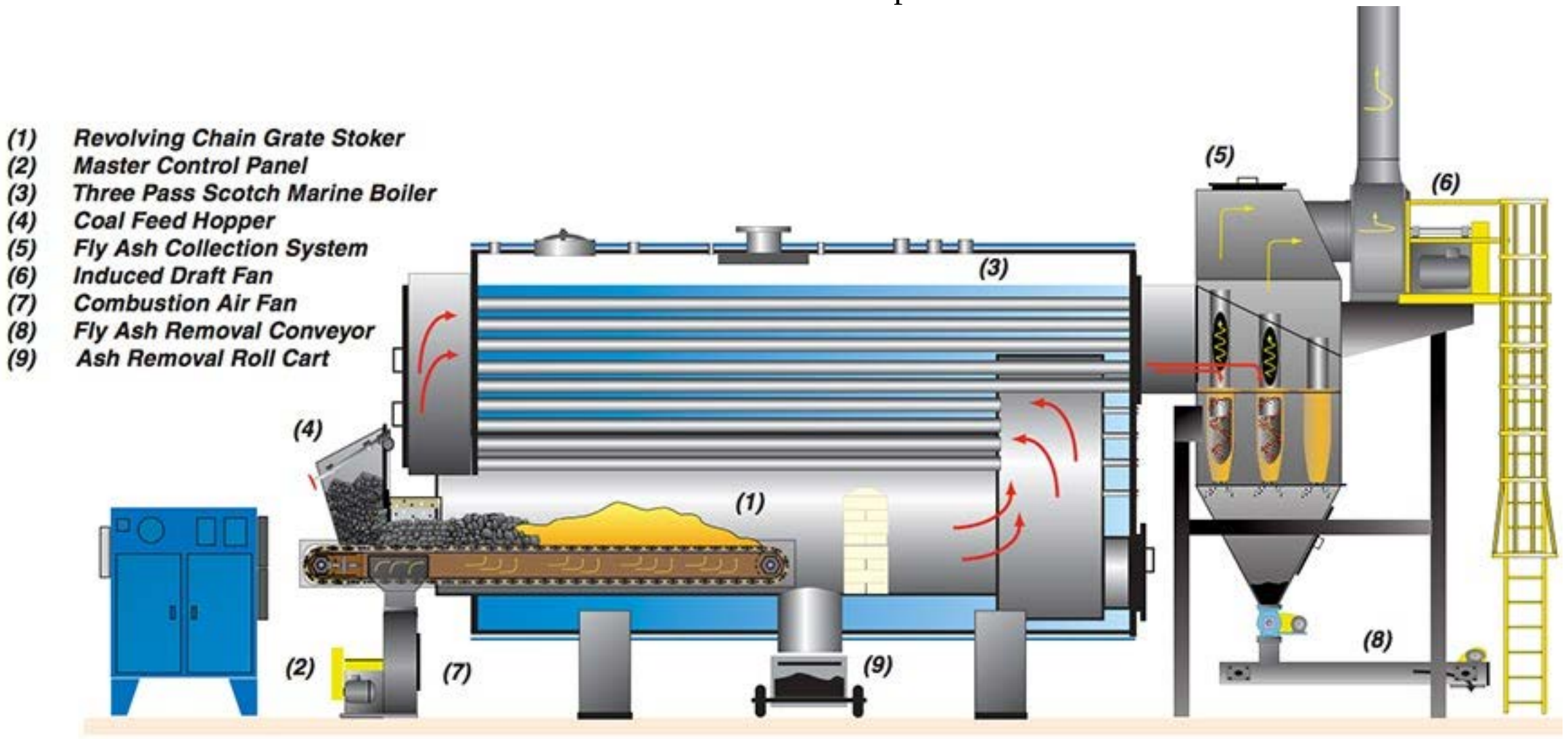

Furneces idustrial dual fuel

2.6 **Natural gas and light Oil furnace:** This type of furnace uses a combination of natural gas and oil to heat the air or other material in the furnace. It is a good choice for areas with variable fuel prices, as it can switch to the more economical fuel source.

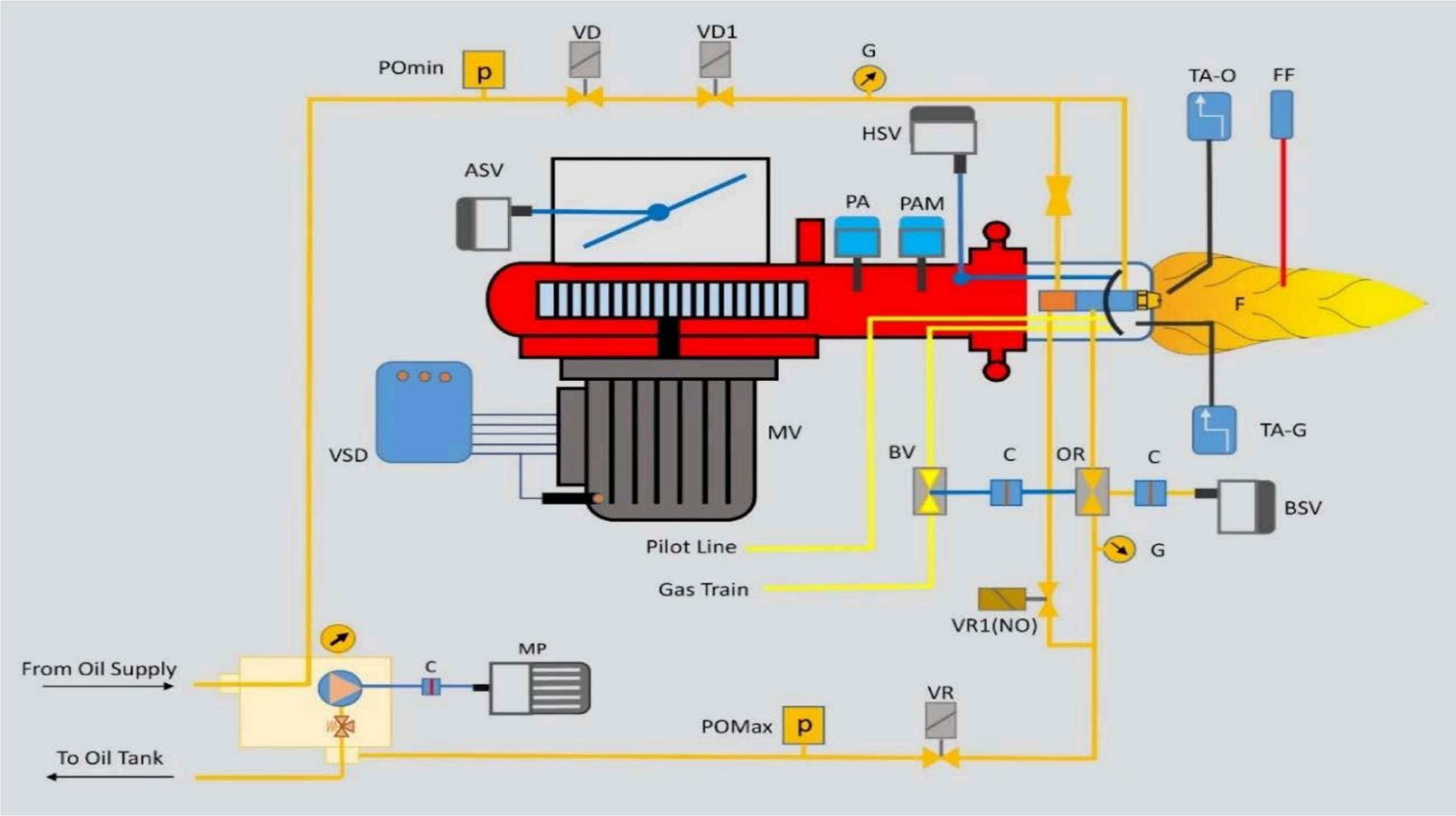

2.7 **Heavy Oil and LPG gas furnace:** This type of furnace uses a combination of oil and gas to heat the air or other material in the furnace. It is a good choice for areas where natural gas is not available, or where the cost of oil is lower than the cost of natural gas.

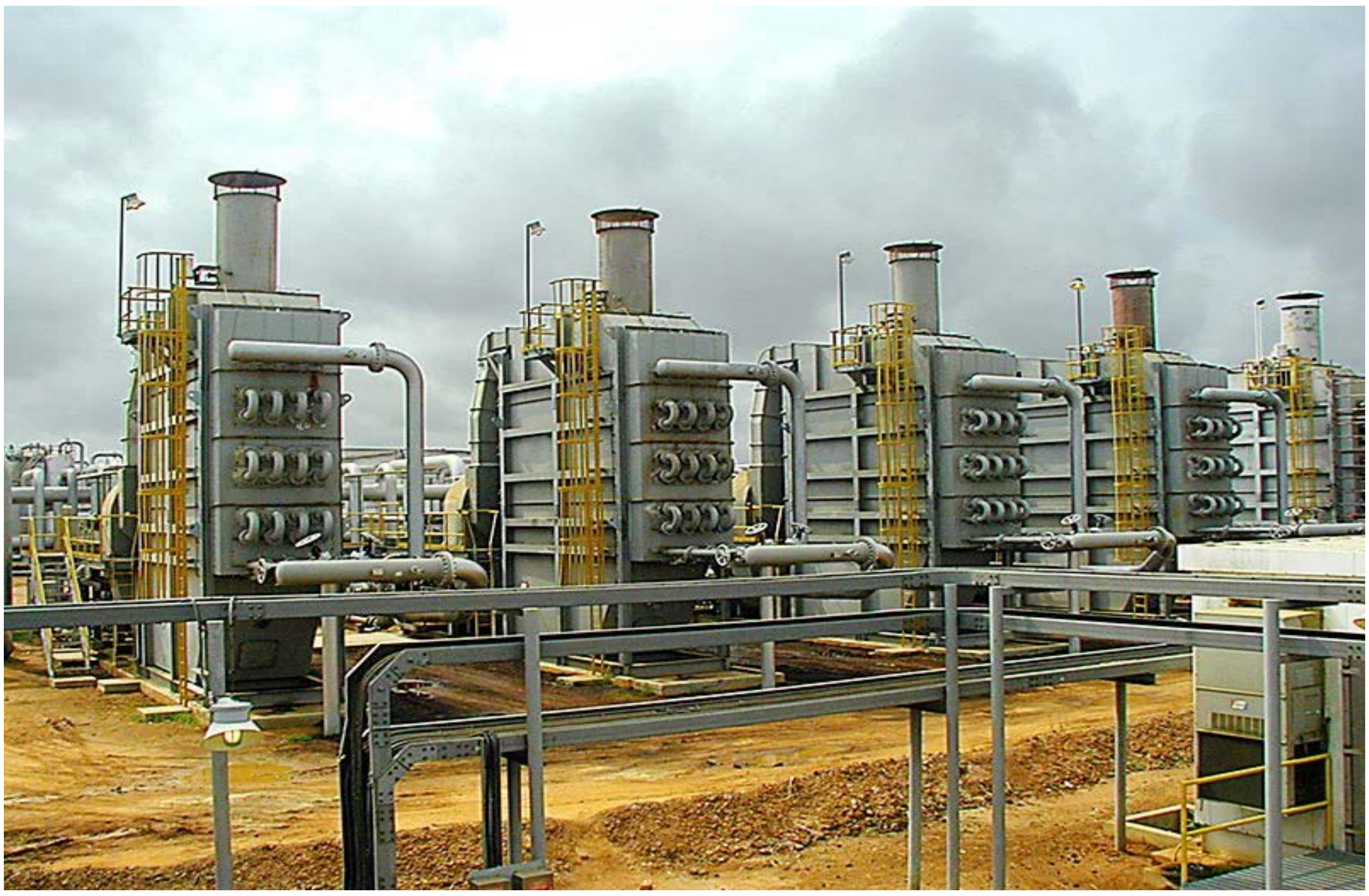

2.8 **Electric and gas furnace:** This type of furnace uses a combination of electricity and gas to heat the air or other material in the furnace. It is a good choice for areas where electricity is less expensive than gas, or where there are environmental regulations that restrict the use of gas.

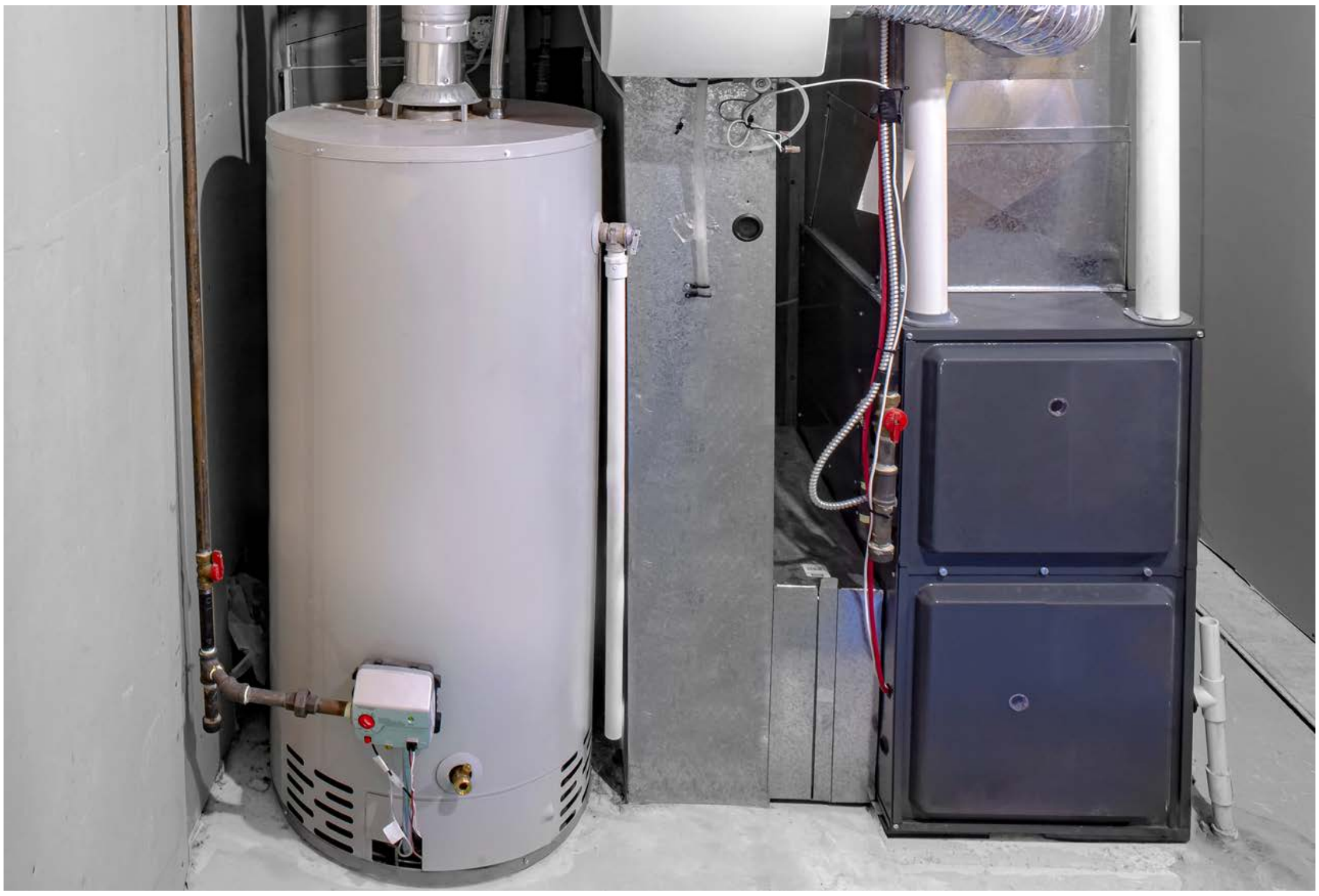

The choice of the best dual fuel furnace for an industrial application will depend on a number of factors, including the climate, the cost of fuels, and the environmental regulations in the area.

Here are some of the benefits of using dual fuel furnaces in industrial applications:

- They can save money on fuel costs, especially if the prices of the two fuels fluctuate.
- They can provide backup heat in case one fuel source is unavailable.
- They can reduce emissions, as they can use the cleaner-burning fuel when available.

However, dual fuel furnaces can also be more expensive to purchase and install than single fuel furnaces. They also require more maintenance and can be more complex to operate.

If you are considering using a dual fuel furnace in your industrial application, it is important to weigh the pros and cons carefully to decide if it is the right choice for you.

The choice of fuel for a furnace also depends on the environmental regulations in the area where the furnace is located. Some fuels, such as wood, Coal may produce more pollutants than others.

## 3. Pros, Cons of Air-To-Fuel Ratio:

The air-to-fuel ratio is the ratio of the mass of air to the mass of fuel in a fuel-air mixture. The stoichiometric air-fuel ratio is the ratio at which the fuel is completely burned with no excess air or fuel. For gasoline, the stoichiometric air-fuel ratio is approximately 14.7:1, which means that for every gram of fuel, 14.7 grams of air are required. [7]

### 3.1 Pros of a stoichiometric air-fuel ratio:

- Complete combustion: Stoichiometric air-fuel ratio ensures complete combustion of the fuel, which produces maximum power and efficiency.
- Fewer emissions: Stoichiometric air-fuel ratio produces fewer emissions, as there is no excess fuel or air to be released into the atmosphere. [9]
- Smoother running: Stoichiometric air-fuel ratio helps to keep the engine running smoothly and prevents knocking or pinging.

### 3.2 Cons of a stoichiometric air-fuel ratio:

- Not always optimal: Stoichiometric air-fuel ratio may not be the optimal ratio for all conditions. For example, a richer mixture may be required for cold starts or high-performance driving. [8]
- Requires precise control: Stoichiometric air-fuel ratio requires precise control of the fuel injection system to ensure that the correct amount of fuel is being injected.
- Can be less fuel-efficient: Stoichiometric air-fuel ratio can be less fuel-efficient than leaner mixtures at low engine loads.

### 3.3 Lean air-fuel ratios:

Lean air-fuel ratios are mixtures with more air than fuel. Lean air-fuel ratios can be used to improve fuel economy and reduce emissions. However, they can also lead to engine knocking and pinging. [8]

### 3.4 Rich air-fuel ratios:

Rich air-fuel ratios are mixtures with more fuel than air. Rich air-fuel ratios can be used to improve power and performance. However, they can also lead to increased emissions and decreased fuel economy. [8]

The optimal air-fuel ratio for a particular engine will depend on a number of factors, including the engine's design, the fuel being used, and the desired performance.

## 4. How to Check Air to Fuel Ratio

There are a few ways to check the air-to-fuel ratio in industrial furnaces. One way is to use a flue gas analyzer. A flue gas analyzer is a device that measures the concentration of various gases in the flue gas, including oxygen, carbon monoxide, and carbon dioxide. The air-to-fuel ratio can then be calculated from these measurements.[11]

Another way to check the air-to-fuel ratio is to use a CO2 probe. A CO2 probe is a sensor that measures the concentration of carbon dioxide in the flue gas. The air-to-fuel ratio can then be calculated from this measurement.

Finally, the air-to-fuel ratio can also be checked by comparing the temperature of the flue gas to the temperature of the furnace. The flue gas will be cooler if there is more air in the mixture, and it will be hotter if there is more fuel in the mixture.

The specific method that is used to check the air-to-fuel ratio will depend on the type of furnace and the availability of equipment.

Here are some additional things to keep in mind when checking the air-to-fuel ratio in an industrial furnace:

- The furnace should be operating at its normal operating conditions.
- The flue gas should be sampled from a representative location in the furnace.
- The flue gas analyzer or CO2 probe should be properly calibrated.

- The results of the air-to-fuel ratio measurement should be interpreted in accordance with the furnace manufacturer's recommendations.

  If the air-to-fuel ratio is not within the desired range, it may be necessary to adjust the fuel system or the combustion air system. The specific adjustments that are needed will depend on the specific problem that is being encountered.

  It is important to note that adjusting the air-to-fuel ratio can be dangerous if done incorrectly. It is best to leave this task to a qualified technician.

  Here's a brief overview of how exhaust gas analysis is conducted and what the measurements indicate:

1. **Oxygen (O2) Measurement:**
   Oxygen levels in the flue gas provide insights into the completeness of combustion. Too much oxygen can indicate excess air, while too little oxygen may suggest incomplete combustion. The optimal oxygen level depends on the type of fuel and combustion conditions.
2. **Carbon Dioxide (CO2) Measurement:**
   CO2 levels are directly related to the efficiency of combustion. High CO2 levels generally indicate efficient combustion, while low levels can suggest incomplete combustion. Monitoring CO2 is essential for assessing the air-to-fuel ratio.
3. **Carbon Monoxide (CO) Measurement:**
   Elevated levels of CO in the flue gas can be an indicator of incomplete combustion. Efficient combustion processes aim to minimize CO emissions, as high levels pose health and environmental risks.
4. **Nitrogen Oxides (NOx) Measurement:**
   NOx compounds, including nitric oxide (NO) and nitrogen dioxide (NO2), are pollutants associated with combustion. Monitoring and controlling NOx emissions are critical for meeting environmental regulations. High combustion temperatures and excess air can contribute to NOx formation.
5. **Sulfur Dioxide (SO2) Measurement:**
   If the fuel being burned contains sulfur, the combustion process can produce sulfur dioxide. Monitoring SO2 emissions is crucial to comply with environmental standards, as sulfur dioxide contributes to air pollution and acid rain.

## 5. How to adjust the air to fuel ratio?

Adjusting the air-to-fuel ratio in industrial furnaces is critical for optimizing combustion efficiency, minimizing emissions, and ensuring proper operation. The specific method for adjusting the ratio can depend on the type of furnace, the fuel used, and the combustion control system in place. Here are general steps that can guide the adjustment process:[1]

1. **Use Combustion Control Systems:**
   - Many industrial furnaces are equipped with combustion control systems that automatically adjust the air-to-fuel ratio based on real-time measurements. These systems often utilize sensors to monitor parameters like oxygen concentration in the flue gas. Adjustments are then made to maintain the desired ratio.
2. **Analyze Flue Gas Composition:**
   - Regularly analyze the composition of the flue gas to understand the combustion process. The presence of certain gases, such as oxygen, carbon dioxide, and carbon monoxide, can provide insights into the combustion efficiency and guide adjustments to the air-to-fuel ratio.
3. **Refer to Combustion Charts and Curves:**
   - Combustion charts and curves specific to the type of fuel being used can provide guidance. These charts typically illustrate the relationship between the air-to-fuel ratio and parameters such as combustion efficiency and emissions. Operators can refer to these charts to identify the optimal ratio.
4. **Adjust Dampers and Air Inlets:**
   - In some cases, manual adjustment of dampers or air inlets may be necessary. This involves regulating the flow of air into the combustion chamber. Care should be taken not to introduce excessive air, which can lead to energy loss and increased emissions.

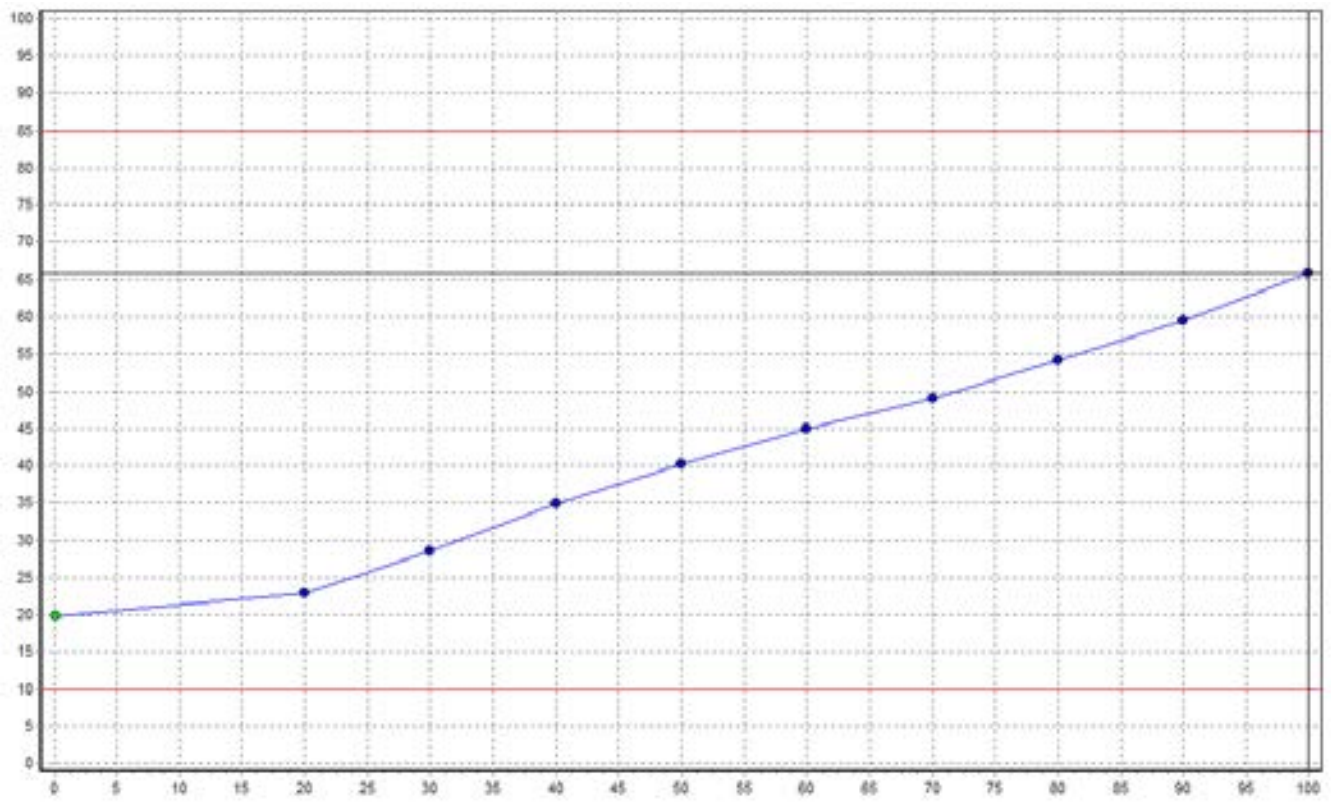

In every furnace we have 4 channels for auxiliary,
Air, Fuel (gas or oil or etc.…), inverter for fan speed
To control the air pressure.

## 6. WHY IS THE AIR TO FUEL RATIO IMPORTANT

In combustion processes, the ratio of air to fuel is typically expressed in terms of mass. The maximum amount of useful heat energy is produced when the mass flow rate of air entering the combustion zone is properly matched to the mass flow rate of fuel entering the burner.[9]

The following generic equation can be used to describe the combustion of fuel:

Fuel + Air → $CO_2$ + $H_2O$ + Heat [9]
where:
$CO_2$ = carbon dioxide
$H_2O$ = water

Air = 21% oxygen ($O_2$) and 79% nitrogen ($N_2$)

Fuel = hydrocarbon such as natural gas or liquid fuel oil

Air is mostly made up of oxygen and nitrogen. It is the oxygen in the air that reacts with the carbon in the fuel to produce carbon dioxide and water vapor. This reaction is highly exothermic, meaning that it releases a large amount of heat energy.

Carbon dioxide is a greenhouse gas, so it is important to ensure that the combustion process is as efficient as possible in order to minimize emissions. Water vapor is also produced during combustion, but it is not a greenhouse gas.[9]

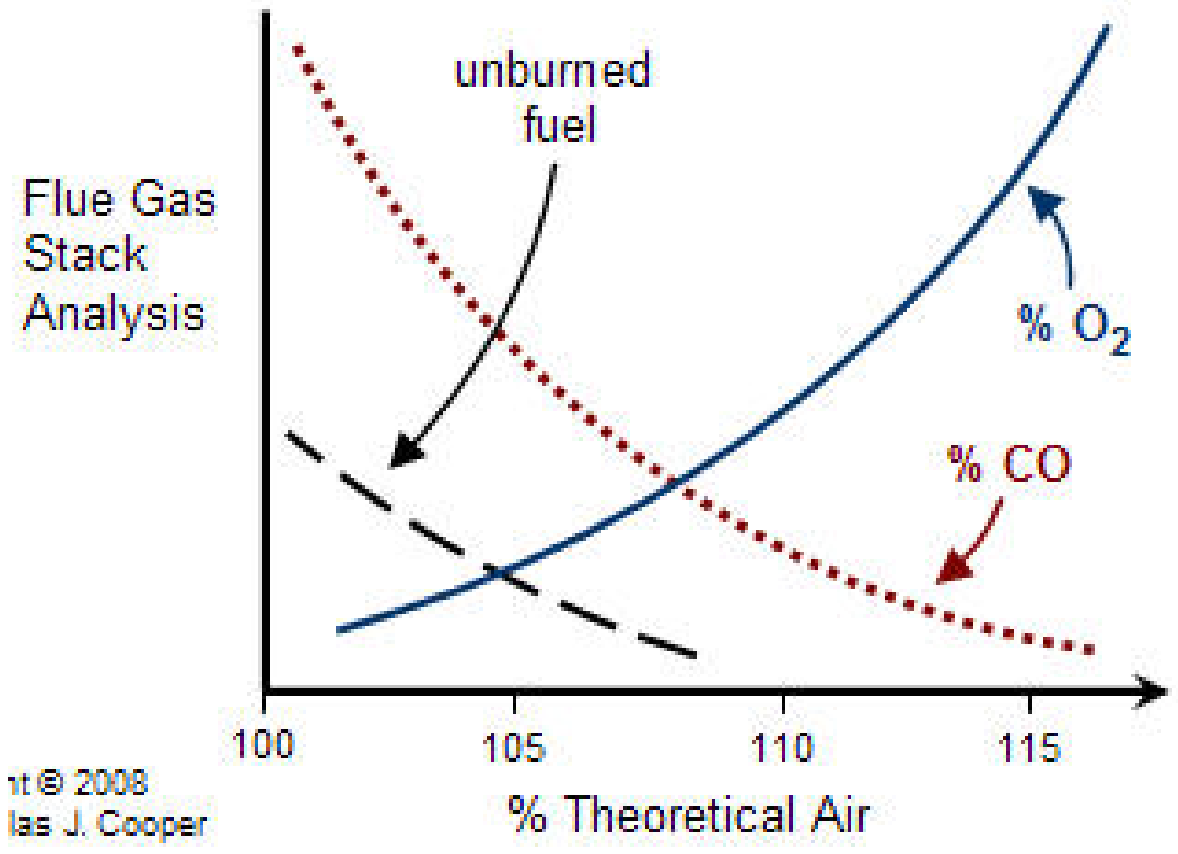

The relationship between the air-to-fuel ratio, pollution formation, and wasted heat energy can be used to design control systems. In a laboratory experiment with perfect mixing and unlimited time, we can determine the minimum amount of air required to completely convert a hydrocarbon fuel to carbon dioxide and water. This minimum amount is called the "stoichiometric" air.

However, real combustion processes do not have perfect mixing or unlimited time. As a result, if we feed air in the exact stoichiometric proportion to the fuel, we will still have incomplete combustion and wasted energy.

Real burners generally perform in a way that is similar to the graph below. The cost associated with operating at increased air-to-fuel ratios is the wasted energy used to heat extra oxygen and nitrogen. However, as the air-to-fuel ratio is decreased, losses due to incomplete combustion and pollution generation increase rapidly.

For any particular burner design, there is a target air-to-fuel ratio that balances the competing effects to minimize the total losses. As the graph above suggests, a gas or liquid fuel burner generally balances losses by operating somewhere between 105% to 120% of theoretical air. This is commonly referred to as operating with 5% to 20% excess air.

## 7. Results:

In these results, the results of different exhaust analysis will be reviewed under different conditions and for burners of different design and operating method adjusting the air/fuel ratio became very useful because it effects on the combustion positively, high efficiency means low fuel consumption, low money, low maintenance, And green environment:

Note: (These results take in different situation different burners fuel (oil & gas), different ambient temperature and different humidity)

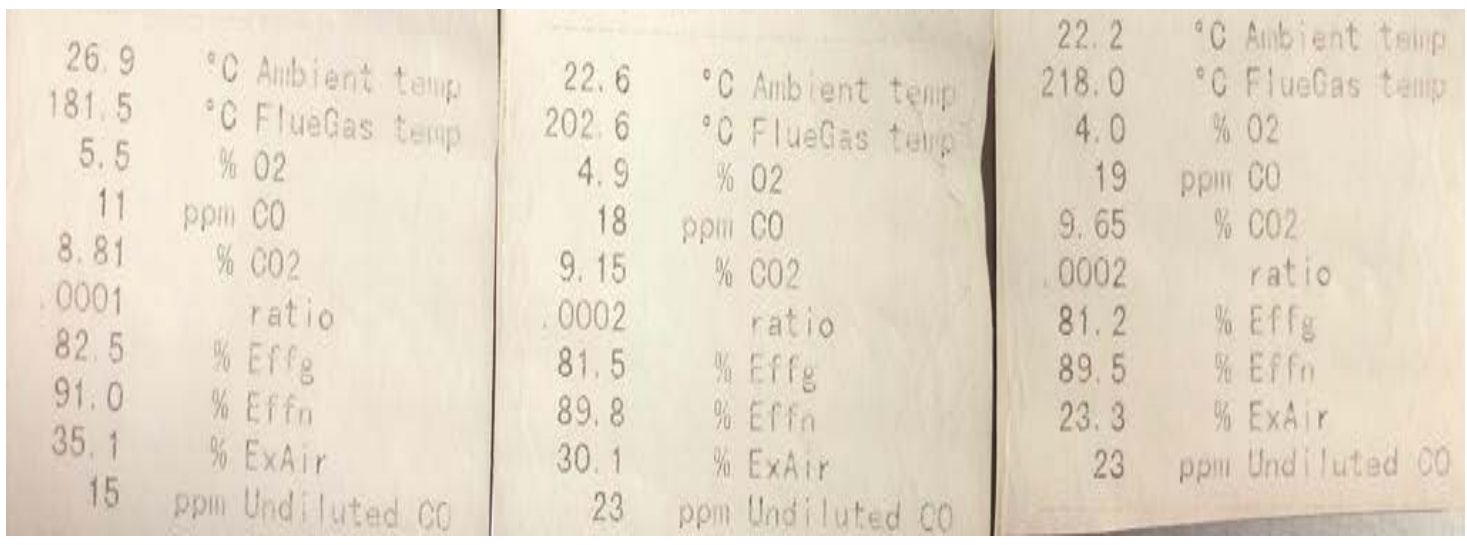

| Variables | Before adjust Air to Fuel ratio | | | After adjust Air to Fuel ratio | | |
|---|---|---|---|---|---|---|
| Ambient temp. | 26.9 | 22.6 | 22.2 | 26.9 | 22.6 | 22.2 |
| Flue Temp. | 219.5 | 232 | 237 | 181.5 | 202 | 218 |
| $O_2$ | 3.1 | 2.9 | 2.1 | 5.5 | 4.5 | 4 |
| CO | 77 | 82 | 106 | 11 | 18 | 19 |
| $CO_2$ | 12.9 | 13.3 | 13.8 | 8.81 | 9.15 | 9.65 |
| Efficiency | 83 | 81 | 78 | 91 | 89 | 89.5 |
| EX Air | 21 | 18.7 | 14.3 | 35 | 30 | 23.3 |

In this report the carbon in high fire rating decreased from 67 ppm to 19 ppm and the efficiency increase from 82% to 91% on low fire and 89% on high fire. (this burner is working with light oil fuel with two servo motors for air and fuel)

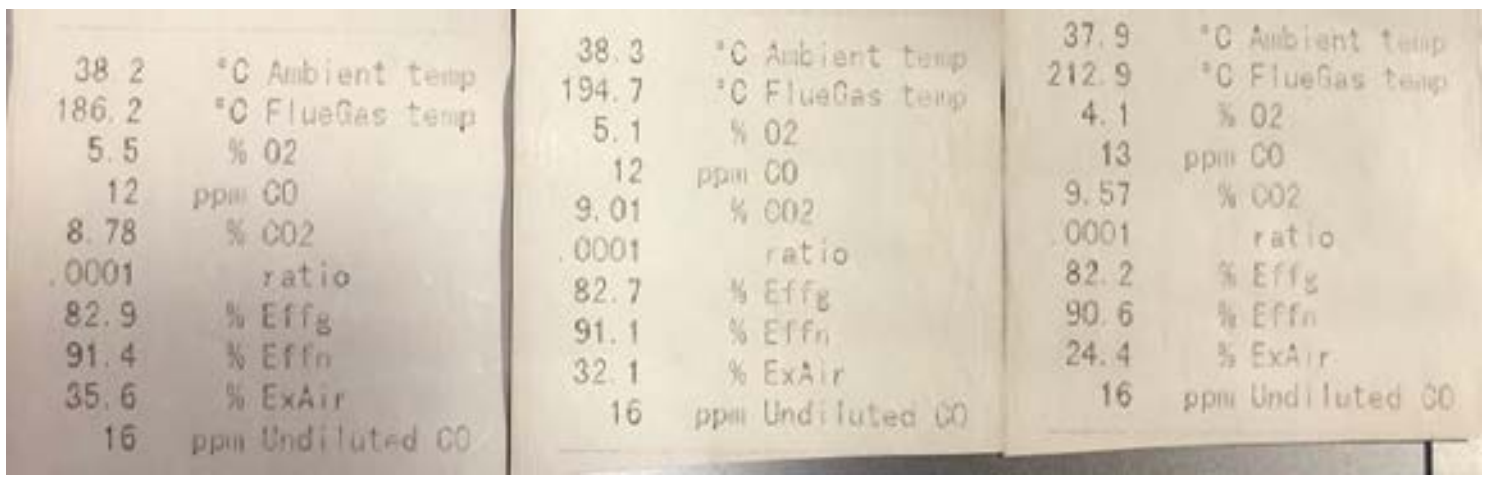

| Variables | Before adjust Air to Fuel ratio | | | After adjust Air to Fuel ratio | | |
|---|---|---|---|---|---|---|
| Ambient temp. | 38 | 38 | 37.9 | 38 | 38 | 37.9 |
| Flue Temp. | 206 | 232 | 251 | 186 | 194 | 212 |
| $O_2$ | 4.2 | 3.9 | 3.2 | 5.5 | 5.1 | 4.1 |
| CO | 46 | 57 | 78 | 12 | 12 | 13 |
| $CO_2$ | 10.78 | 11.01 | 11.57 | 8.78 | 9.01 | 9.57 |
| Efficiency | 86.4 | 83.1 | 80.6 | 91.4 | 91.1 | 90.6 |
| EX Air | 24.6 | 22 | 19 | 35.6 | 32 | 24.4 |

In this report the carbon in high fire rating decreased from 54 ppm 13 ppm and the efficiency increase from 84.2 % to 91.4% on low fire and 90.6% on high fire. (this burner is working with natural gas fuel with three servo motors for air ,fuel and AUX)

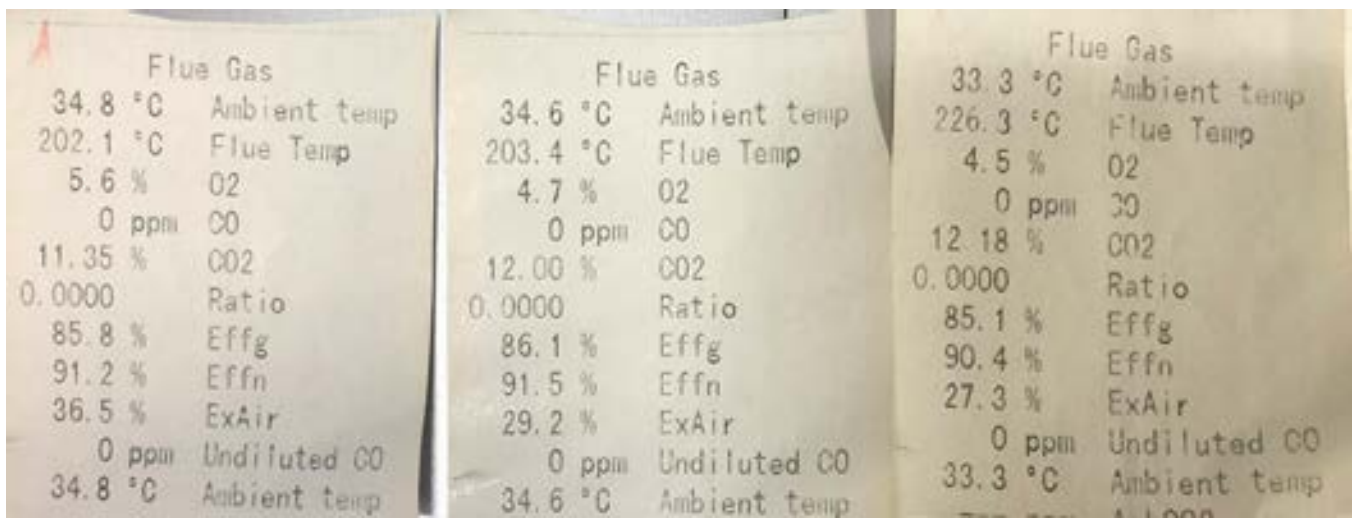

| Variables | Before adjust Air to Fuel ratio | | | After adjust Air to Fuel ratio | | |
|---|---|---|---|---|---|---|
| Ambient temp. | 26.9 | 22.6 | 22.2 | 34 | 34 | 33 |
| Flue Temp. | 219.5 | 232 | 237 | 202 | 203 | 226 |
| $O_2$ | 3.1 | 2.9 | 2.1 | 5.6 | 4.7 | 4.5 |
| CO | 77 | 82 | 106 | 0 | 0 | 0 |
| $CO_2$ | 12.9 | 13.3 | 13.8 | 8.81 | 9.15 | 9.65 |
| Efficiency | 83 | 81 | 78 | 91 | 89 | 89.5 |
| EX Air | 21 | 18.7 | 14.3 | 35 | 30 | 23.3 |

In this case the CO always zero (this is special case) this burner is low NOx Burner and the efficiency between 91.2% on low fire and 90.4% on high fire. (this burner is low NOx Burner natural gas fuel and this burner with AUX control and with VSD for air control)

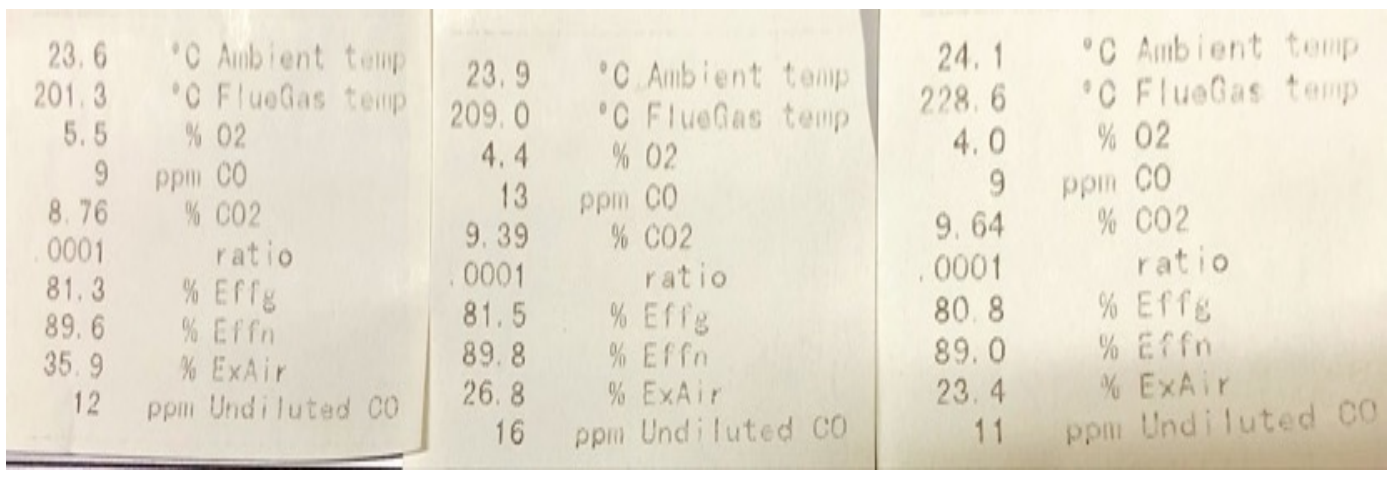

| Variables | Before adjust Air to Fuel ratio | | | After adjust Air to Fuel ratio | | |
|---|---|---|---|---|---|---|
| Ambient temp. | 23.6 | 23.9 | 24.1 | 23.6 | 23.9 | 24.1 |
| Flue Temp. | 223 | 237 | 241 | 201 | 209 | 228 |
| $O_2$ | 4.7 | 4.1 | 3.2 | 5.5 | 4.4 | 4 |
| CO | 43 | 49 | 66 | 9 | 13 | 9 |
| $CO_2$ | 10.1 | 10.9 | 11.9 | 8.76 | 9.39 | 9.64 |
| Efficiency | 84 | 82.8 | 81 | 89 | 89.8 | 89 |
| EX Air | 27 | 20.5 | 17.4 | 35 | 26 | 23.4 |

In this report the carbon in high fire rating decreased from 61 ppm 13 ppm and the efficiency increase from 86% to 89.6% on low fire and 89% on high fire. (this burner is working with LPG fuel with two servo motors for air and fuel)

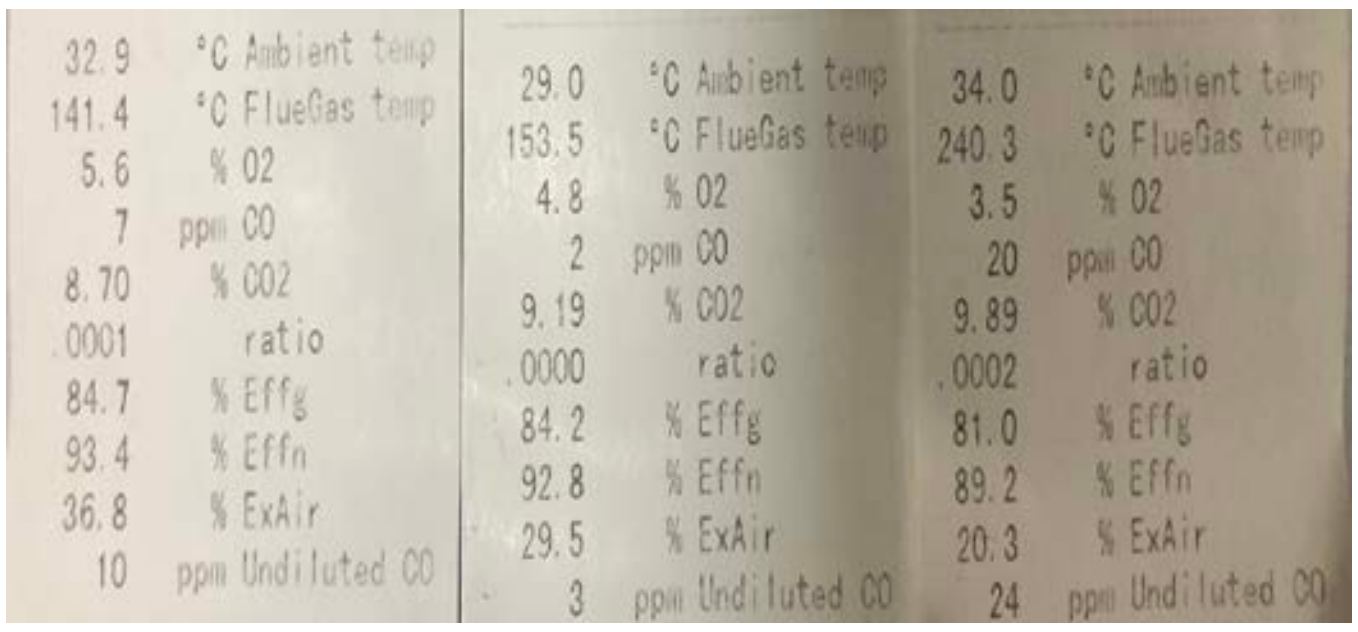

| Variables | Before adjust Air to Fuel ratio | | | After adjust Air to Fuel ratio | | |
|---|---|---|---|---|---|---|
| Ambient temp. | 32.6 | 33.7 | 34 | 32 | 29 | 34 |
| Flue Temp. | 201.5 | 206 | 230 | 141.5 | 153 | 240 |
| $O_2$ | 7.6 | 6.9 | 5.8 | 5.6 | 4.8 | 3.5 |
| CO | 18 | 36 | 31 | 7 | 2 | 20 |
| $CO_2$ | 7.70 | 8.16 | 9.1 | 8.70 | 9.19 | 9.89 |
| Efficiency | 82.4 | 83.1 | 80.6 | 93.4 | 92.5 | 89.2 |
| EX Air | 45 | 39 | 32 | 36 | 29.5 | 20.3 |

In this report the carbon in high fire rating decreased from 73 ppm 20 ppm and the efficiency increase from 87% to 93.4% on low fire and 89.2% on high fire. (this burner is working with dual fuel light oil fuel & LPG with three servo motors for air and fuel and VSD for rotary cup)

## 8. CONCLUSION:

The air-to-fuel ratio is an important factor in combustion processes. The optimal air-to-fuel ratio for a particular burner design will depend on a number of factors, including the type of fuel being used, the desired performance, and the environmental regulations.

In general, a stoichiometric air-to-fuel ratio is the minimum amount of air required to completely convert a hydrocarbon fuel to carbon dioxide and water. However, real combustion processes do not have perfect mixing or unlimited time, so operating at the stoichiometric air-to-fuel ratio can lead to incomplete combustion and wasted energy.

As a result, most burners operate with a slightly richer air-to-fuel ratio, typically between 105% and 120% of theoretical air. This is commonly referred to as operating with 5% to 20% excess air. This helps to ensure complete combustion and minimize pollution emissions.

The air-to-fuel ratio can be adjusted by controlling the amount of air or fuel entering the burner. This can be done manually or automatically using a control system.

The air-to-fuel ratio is an important factor in the efficiency, emissions, and performance of combustion processes. By carefully controlling the air-to-fuel ratio, it is possible to improve these factors and minimize the environmental impact of combustion processes.